\documentclass{appolb}
\usepackage{epsfig}

\begin{document}
\title{STOCHASTIC NETWORK VIEW ON HADRON PRODUCTION}
\author{G.Wilk\thanks{e-mail: wilk@fuw.edu.pl}
\address{The Andrzej So\l tan Institute of Nuclear Studies, 
         Ho\.za 69; 00-689 Warsaw, Poland}
\and  Z. W\l odarczyk\thanks{wlod@pu.kielce.pl}
\address{Institute of Physics, \'Swi\c{e}tokrzyska Academy,
         \'Swi\c{e}tokrzyska 15; 25-406 Kielce, Poland
}
}
\maketitle

\begin{abstract}
We demonstrate that hadron production viewed as formation of specific
stochastic network can explain in natural way the power-law
distributions of transverse mass spectra of pions found recently,
which seem to substitute the expected Boltzmann statistical factor. 
\\    

\end{abstract}

Recently it has been pointed out \cite{MGMG} that properly normalized
transverse mass spectra of $\pi^0$ mesons for $m_T > 1$ GeV/c$^2$
obey specific $m_T$ scaling, namely they follow the universal
power-law curve, 
\begin{equation}
\frac{dN}{m_T^2 dm_T}\, =\,  c\cdot
\left(\frac{m_T}{\Lambda}\right)^{-P}, \label{eq:mt} 
\end{equation}
(which after integration over transverse momenta results in similar
power law in masses and, according to \cite{MGMG}, describes well
the yields of neutral mesons from $\eta$ to $\Upsilon$, i.e., for
$m\simeq 0.5-10$ GeV/c$^2$) with $P\sim 8-10$, depending on energy
and type of particles. This has been regarded as purely
phenomenological observation of the apparent violation of the usual
Boltzmann behaviour, 
\begin{equation}
\frac{dN}{m_T^2 dm_T}\, =\,  c_{BG}\cdot \exp\left(
-\frac{m_T}{\Lambda}\right), \label{eq:bgmt}  
\end{equation}
in the region of high $m_T$. In \cite{WWq} we have noticed that this
finding can be regarded also as confirmation that in descibing
hadronization process one should apply not the usual Boltzmann-Gibbs
(BG) statistics but rather its nonextensive generalization, for
example in the form of the Tsallis statistics \cite{T}, in which eq.
(\ref{eq:bgmt}) is replaced by 
\begin{equation}
\frac{dN}{m_T^2 dm_T}\, =\,  c_q\left[ 1 -
(1-q)\frac{m_T}{\Lambda}\right]^{\frac{1}{1-q}}. \label{eq:qmt} 
\end{equation}
Obviously, for large values of $m_T$ (where dependence to scale
$\Lambda$ can be neglected) eq. (\ref{eq:qmt}) becomes eq.
(\ref{eq:mt}). However, contrary to eq. (\ref{eq:mt}) the above
distribution offers simple interpretation of the fitted parameters,
$q$ and $\Lambda$. In particular, as demonstrated in \cite{WW}
\footnote{We refer interested reader to \cite{WW,T} for details and
further references.}, $q$ is entirely given by fluctuations of
parameter  $1/\Lambda$ in the usual BG approach, i.e., in eq.
(\ref{eq:bgmt}) \footnote{Notice that usually $\Lambda = T$, i.e., it
is regarded as the "temperature" of the hadronizing system treated as
a kind of "heat bath". Fluctuations could arise, for example, from
the fact that, in cases considered here, such heat bath is in obvious
way finite \cite{Almeida}.}, with nonextensivity parameter $q=1+1/P$.
Notice that for $q \rightarrow 1$ Tsallis statistics becomes the
usual BG  one and eq. (\ref{eq:qmt}) becomes eq. (\ref{eq:bgmt}).\\ 

This, however, does not solve the puzzle noticed by \cite{MGMG},
namely what is dynamical origin leading to the apparent scale-free
character of the observed $m_T$ spectra seen in eq. (\ref{eq:mt})
\footnote{Notice that scale parameter $\Lambda$ in eq. (\ref{eq:mt})
can be absorbed in the normalization constant, as it was done in
\cite{MGMG}.}. We would like to propose here one possible scenario
resulting in eq. (\ref{eq:mt}). To this aim let us first mention that
Tsallis distribution (\ref{eq:qmt}) describes also very well the
formation of the so called complex free networks \cite{WWNET} (if one
replace $m_T$ by the number of links $k$)\footnote{Actually, in
\cite{WWNET} the so called escort probability distribution was used,
which results in $(\dots)^{\frac{q}{1-q}}$ instead of
$(\dots)^{\frac{1}{1-q}}$ as here. This has no consequences in what
concerns our work as both $q$ can be simply translated to each
other.}. Inspired by this observation we shall investigate in what
follows the possibility that power law seen in eq. (\ref{eq:qmt})
signals that hadronization can be viewed (at least from the limited
perspective considered here) as process of formation of some specific
network taking place in the environment of gluons and quark-antiquark
pairs $(q\bar{q})$, process in which their original actual
energy-momentum distributions would be of second importance in
comparison to the fact that, because of their mutual interactions,
they connect to each other and that this process of connection has
its distinctive dynamical consequences.\\  

The proposed line or reasoning is following. Suppose that we start
with some initial state consisting of number $n_0$ of already existed
$(q\bar{q})$ pairs, which we shall consider as equivalent to vertices
in the usual network. We shall now add to them, in each consecutive
time step, another vertex (i.e., a new $(q\bar{q})$ pair), which can
have $k_0$ possible connections to the old state. To be more specific,
we regard our quarks as dressed by interaction with surrounding
gluons and therefore somehow "excited". Each quarks is supposed to
interact with $k$ other quarks, what in network terminology would
mean that each quark has $k$ links. We shall now assume that the
"excitation" of quark mentioned above is proportional to $k$. We
expect also that the number of links $k$ should be proportional to
the number of gluons participating in such "excitation", i.e.,
existing in the vicinity of a given quark. In this case the natural
consequence would be that the chances to interact with a given quark
grow with the number of links $k$ attached to it and are equal to
\begin{equation}
w(k_i)\, =\, \frac{k_i}{\sum k_i} , \label{eq:chances}
\end{equation}
i.e., that new links will be preferentially attached to quarks with
already large values of $k$. This corresponds to building up of the
so called preferential network, which evolves due to the occurence of
new $(q\bar{q})$ pairs from decaying gluons. Because of this the
number of links is here twice the number of links in the usual
network (cf. \cite{AB}). After time $t$ one has therefore  
\begin{equation} 
\sum k_i = 2(2k_0t) = 4k_0t \label{eq:links}
\end{equation}
links. This leads to the following growth equation for the $i-$th
object (the $i-$th $(q\bar{q})$ pair):
\begin{equation}
\frac{\partial k_i}{k_i} = \frac{1}{\delta}\cdot \frac{\partial
t}{t}, \label{eq:growth} 
\end{equation}
where $\delta = 4$. Solving this equation for the initial condition
stating that the $i-$th object appears in time $t_i$ with the number
of links equal to $k_0$ one gets that
\begin{equation}
k_i(t) = k_0 \left(\frac{t}{t_i}\right)^{\frac{1}{\delta}} .
\label{eq:growtha} 
\end{equation}
Notice that probability of forming $k_i(t) < k$ links is then given
by 
\begin{equation}
P(k_i(t)<k) = P\left(t_i > \frac{k_i^{\delta}\cdot
t}{k_0^{\delta}}\right). \label{eq:der} 
\end{equation}
Assuming now uniform probability distribution of occurence of objects
$(q\bar{q})$, the probability of adding to our system a new such
object in the unit of time is given by 
\begin{equation}
P(t_i) = \frac{1}{t} . \label{eq:Pt}
\end{equation}
It means therefore that, because of eq. (\ref{eq:der}),
\begin{equation}
P(k_i(t)<k) = P\left(t_i > \frac{k_i^{\delta}\cdot
t}{k_0^{\delta}}\right) = 1 - P\left(t_i\leq
\frac{k_0^{\delta}}{k^{\delta}}\cdot t \right) = 1 -
\left(\frac{k_0}{k}\right)^{\delta} \label{eq:deriv}  
\end{equation}
and probability of forming object with $k$ links can be written as
\begin{equation}
P(k) = \frac{\partial P(k_i(t)<k)}{\partial k} =
\frac{k_0^{\delta}}{1+\delta}\cdot k^{-(1+\delta)}, \label{eq:derivat}
\end{equation}
i.e., the resulting distribution of the number of links is
independent of time and has the characteristic power-like form:
\begin{equation}
P(k) \sim k^{-\gamma} \label{eq:pk}
\end{equation}
with $\gamma = 1+\delta$ (for (\ref{eq:links}) $\gamma = 5$). For the
general case of \footnote{It accounts for the fact that new vertex
$(q\bar{q})$ arises from the $gluon \rightarrow q\bar{q}$ process,
i.e., one gluon disapears and this corresponds to diminishing the
number of links by one for each two vertices.} 
\begin{equation}
\sum k_i = 2t(2k_0-1)  \label{eq:growth1} 
\end{equation}
one has $\delta = \frac{2(2k_0-1)}{k_0} = 4 - \frac{2}{k_0}$, i.e., the
same eq. (\ref{eq:pk}) but with $\gamma = 5 - \frac{2}{k_0}$
\footnote{Notice that in the approach to networks using Tsallis
statistics \cite{WWNET} one gets $P(k) \sim k^{\frac{q}{1-q}}$, i.e.,
$\gamma =5$ corresponds to $q=1.25$.}.\\  

The crucial point is now our conjecture that the number of links $k$
determines the transverse mass of emitted particle $m_T$. Assuming
that 
\begin{equation}
m_T \sim k^{\alpha}  \label{eq:x}
\end{equation}
one gets immediately that 
\begin{equation}
P(m_T) \sim m_T^{-\frac{\gamma}{\alpha}}\cdot
m_T^{\frac{1-\alpha}{\alpha}} = m_T^{-\beta}\qquad {\rm where}\qquad
\beta = 1 + \frac{\gamma -1}{\alpha} . \label{eq:Px}
\end{equation}
This completes our derivation of (\ref{eq:mt}) where now $P= \beta$.
It arises because of the conjecture (\ref{eq:x}) connecting of the
actual value of $m_T$ with the number of links in some directional
network growing process defined above. The first parameter here is
$k_0$, which denotes the number of links with which the new vertex
(here the $(q\bar{q})$ pair) will join the already existing network
of such vertices. However, as one can see, with increasing $k_0$ one
quickly obtains asymptotic value of $\gamma = 5$. The other
parameter, $\alpha$, describes in what way the new links can be
chosen in momentum spce in respect to allowed directions. For
example, for totally random distribution one should choose,
analogously to deterministic diffusion process, $\alpha =
\frac{1}{2}$ \footnote{Deterministic diffusion concept is widely
discused in \cite{DD}. Notice that lack of such diffusion in momentum
space would mean that interactions are highly correlated, its presence
indicates chaotic evolution in momentum space.}. Other values of
parameter $\alpha$ would correspond to other special diffusion cases
(see below). In case of $\alpha = \frac{1}{2}$ one has $\beta = 9$
for eq. (\ref{eq:links}) and $\beta = 9-\frac{4}{k_0}$ for eq.
(\ref{eq:growth1}), which numerically gives us values of $P$ equal to
$7.66,~8.0$ and $8.5$ for, respectively, $k_0=3,4$ and $8$. As one
can see, they agree with those quoted by \cite{MGMG} (and listed
before).\\ 

It is worth to stress that proposed approach leads naturally to
distinct possible behaviours of transverse mass distributions (all
others will be their suitable combination):  
\begin{itemize}
\item[$(a)$] The power-like distribution of the type given by eq.
(\ref{eq:pk}) (which was our main motivation). It corresponds to the
case of large excitations for which probability of connecting a new
quark to the one already existing in the system depends on the number
of the actual connections realised so far. Large number of connection
results then in large excitation, what means that one has large
emission of gluons and what, finally, enhances chances of connection
to such a quark. 
\item[$(b)$] If one assumes instead that the new quarks attaches
itself to the already existing one with equal probability (such would
be situation for small excitations, i.e., for small $p_T$) one gets
instead exponential distribution of links,
\begin{equation}
P(k) \sim \exp\left(-\frac{k}{\langle k\rangle}\right), \label{eq:expk}
\end{equation}
which results in different distributions of $m_T$ depending on the
type of diffusion given by parameter $\alpha$ in eq. (\ref{eq:x}) and
ranging from
\begin{equation}
P(m_T) \sim \exp\left( - \frac{m_T^2}{\langle m_T^2\rangle}\right)\qquad
{\rm for}\qquad \alpha = \frac{1}{2} \label{eq:exp2} 
\end{equation}
when the full fledged diffusion is allowed to
\begin{equation}
P(m_T) \sim \exp\left( - \frac{m_T}{\langle m_T\rangle }\right) \qquad
{\rm for}\qquad \alpha = 1 . \label{eq:exp} 
\end{equation}
when there is no diffusion. This would be the case of quarks located
on the periphery of the region of hadronization in which case they
could interact only in interior quarks and in such case $m_T \sim k$.
\end{itemize}

To summarize - prompted by the observation of the power-like
behaviour of the transverse mass spectra reported by \cite{MGMG} we
have considered here possibility that they can be reflection not so
much of any special kind of equilibrium (resulting in some specific
statistics) but rather of the formation process, which follows
(directed) free networks formation pattern discussed widely in the
literature and found in many branches of sciences \cite{AB,WWNET}.
Identify vertices in such network as $(q\bar{q})$ pairs and gluons as
links we were able to derive eq. (\ref{eq:mt}), obtained in
\cite{MGMG} by analysis of some experimental data, when we assume that
observed $m_T$ reflects somehow the number of links in such network
(eq. (\ref{eq:x})). The resulting power-like distribution is rather
universal (see, for example \cite{SOC} where similar power-like
distributions observed in other branches of high energy phenomenology
have been atributed to the apparent self organizing character of the
corresponding processes) and its power index depends mainly on the
type of the allowed diffusion process as given by parameter
$\alpha$. The type of this deterministic diffusion process in the
momentum space which is allowed in a given experimental scenario
leads then to three kinds of distinct characteristic spectra of $m_T$
out of which all observed spectra could be composed. \\

Partial support of the Polish State Committee for Scientific Research
(KBN) (grant 2P03B04123 (ZW) and grants
621/E-78/SPUB/CERN/P-03/DZ4/99 and 3P03B05724 (GW)) is acknowledged.\\



\begin{thebibliography}{99}
 
 \bibitem{MGMG} M.Ga\'zdzicki and M.I.Gorenstein, {\sl Phys. Lett.}
                {\bf 517} (2001) 250.  
 
 \bibitem{WWq} G.Wilk and Z.W\l odarczyk, {\sl Physica} {\bf A305}
               (2002) 227.

 \bibitem{T} C.Tsallis, {\sl J. Stat. Phys.} {\bf 52} (1988) 479;
             cf. also C.Tsallis, {\sl Chaos, Solitons and Fractals}
             {\bf 13} (2002) 371, {\sl Physica} {\bf A305} (2002) 1
             and in {\it Nonextensive Statistical Mechanics and its
             Applications}, S.Abe and Y.Okamoto (Eds.), Lecture Notes
             in Physics LPN560, Springer (2000). For updated
             bibliography on this subject see 
             http://tsallis.cat.cbpf.br/biblio.htm. Its recent 
             summary is provided in the special issue of {\sl Braz. 
             J. Phys.} {\bf 29} (No 1) (1999) (available also at 
             http://sbf.if.usp.br/ WWW\_pages/ Journals/BJP/ Vol29/Num1/index.htm). 
             
 \bibitem{WW} G.Wilk and Z.W\l odarczyk, {\sl Phys. Rev. Lett.}
              {\bf 84} (2000) 2770, {\sl Chaos, Solitons and
              Fractals} {\bf 13/3} (2001).

 \bibitem{Almeida} M.Baranger, {\sl Physica} {\bf A305} (2002) 27;
                   M.P.Almeida, {\sl Physica} {\bf A325} (2003) 426.

 \bibitem{WWNET} G.Wilk and Z.W\l odarczyk, {\sl Acta Phys. Polon.}
                 {\bf B35} (2004) 871.
 
 \bibitem{AB} R.Albert and A.L.Barabasi, {\sl Rev. Mod. Phys.} {\bf 74}
              (2002) 47.

 \bibitem{DD} H.Haken, {\it Synergetics}, Springer, Berlin (1982);
              S.Grossman, in {\it Evolution of Order and Chaos}
              (ed. H.Haken), Springer, Berlin (1982); T.Geisel and
              T.Nierwetberg, {\sl Phys. Rev. Lett.} {\bf 48} (1982)
              7. 

 \bibitem{SOC} C.Boros, Meng Ta-chung, R.Rittel, K.Tabelow and Zhang
               Yang, {\sl Phys. Rev.} {\bf D61} (2000) 094010 and
               references therein; Fu Jinghua, Meng Ta-chung,
               R.Rittel and K.Tabelow, {\sl Phys.Rev.Lett.} {\bf 86}
               (2001) 1961.  

\end{thebibliography}
\end{document}